\documentclass{article}
\usepackage{spconf,amsmath,graphicx,hyperref}

\usepackage{multirow}
\usepackage{cite}
\usepackage{mciteplus}
\usepackage{etoolbox}
\usepackage{adjustbox}
\usepackage{booktabs}
\usepackage{float}
\usepackage{subcaption}   
\usepackage{amssymb,amsfonts} 
\usepackage{relsize}
\usepackage{caption} 

\title{RFM-Editing: Rectified Flow Matching for \\ Text-guided Audio Editing}
\vspace{-2mm}

\name{Liting Gao$^{1}$, Yi Yuan$^{1}$, Yaru Chen$^{1}$, Yuelan Cheng$^{1}$, Zhenbo Li$^{2}$, Juan Wen$^{2}$, Shubin Zhang$^{3}$,
Wenwu Wang$^{1}$ \thanks{Project page: \href{https://katelin-glt.github.io/RFM-Editing-Demo/} {https://katelin-glt.github.io/RFM-Editing-Demo/}}}

\address{%
$^{1}$ Centre for Vision, Speech and Signal Processing (CVSSP), University of Surrey, United Kingdom \\
$^{2}$ College of Information and Electrical Engineering, China Agricultural University, China \\
$^{3}$ Fisheries College, Ocean University of China, China
}

\begin{document}

\maketitle

\begin{abstract}
Diffusion models have shown remarkable progress in text-to-audio generation. However, text-guided audio editing remains in its early stages. This task focuses on modifying the target content within an audio signal while preserving the rest, thus demanding precise localization and faithful editing according to the text prompt. Existing training-based and zero-shot methods that rely on full-caption or costly optimization often struggle with complex editing or lack practicality. In this work, we propose a novel end-to-end efficient rectified flow matching-based diffusion framework for audio editing, and construct a dataset featuring overlapping multi-event audio to support training and benchmarking in complex scenarios. Experiments show that our model achieves faithful semantic alignment without requiring auxiliary captions or masks, while maintaining competitive editing quality across metrics.

\end{abstract}


\vspace{-2mm}
\section{Introduction}
\label{sec:intro}
\vspace{-1mm}
Recent advances in diffusion-based modeling have led to remarkable progress in text-to-audio (TTA) generation, with examples including denoising diffusion probabilistic model (DDPM)~\cite{ho2020denoising} based methods (e.g., AudioLDM \cite{liu2023audioldm, liu2024audioldm}, Make-An-Audio \cite{huang2023make, huang2023make2}) and flow \cite{lipman2022flow} based methods (e.g., TangoFlux \cite{hung2024tangoflux}). Text-guided audio editing aims to modify existing audio based on natural language instructions or target descriptions while preserving the unaltered content. This enables flexible audio manipulation through prompts and supports applications in sound design, post-production, and personalized audio generation. However, research on text-guided audio editing, including training-free diffusion-inversion methods \cite{jia2025audioeditor, xu2024prompt, manor2024zero} and training-based models \cite{wang2023audit, paissan2024audio}, remains limited in performance and at an early stage.  

Training-free audio editing typically leverages pre-trained TTA diffusion models \cite{xue2024auffusion, liu2023audioldm}, inverting diffusion to recover latent noise from input audio and guiding denoising with textual prompts. Methods such as AudioEditor \cite{jia2025audioeditor} use denoising diffusion implicit model (DDIM) inversion and null-text optimization for high-fidelity edits, while prompt-guided precise audio editing (PPAE) \cite{xu2024prompt} and DDPM inversion Zero-Shot \cite{manor2024zero} manipulate cross-attention maps for localized control via semantic shifts between source and target prompts. They all introduce the Prompt-to-Prompt attention replacement mechanism \cite{hertzprompt} into audio editing to accurately align with the target text and significantly improve the CLAP score. WavCraft \cite{liang2024wavcraft} further extends this paradigm by using large language models (LLMs) to translate prompts into expert-module instructions for flexible editing.

By contrast, AUDIT \cite{wang2023audit} trains a latent diffusion model (LDM) \cite{rombach2022high} with triplet data for instruction-guided audio editing. Non-rigid prompt editing \cite{paissan2024audio} fine-tunes a diffusion model on audio-caption pairs and performs edits via interpolation in prompt embedding space. 
Although training-based methods \cite{wang2023audit} enable instruction following through explicit supervision, their progress is limited by the scarcity of large-scale datasets, making it difficult to accurately localize edit regions while preserving the rest, especially in complex scenarios with overlapping sounds. In contrast, training-free methods offer flexibility without labeled data, but often require costly null-text optimization in inference \cite{jia2025audioeditor}. Furthermore, some methods depend on full captions \cite{jia2025audioeditor, xu2024prompt, manor2024zero, paissan2024audio} or modified token masks \cite{jia2025audioeditor} rather than concise editing instructions, which is time-consuming and impractical. Since audio is typically not accompanied by detailed textual descriptions, we argue that an ideal audio editing system should operate from raw audio and editing instructions, as in \cite{wang2023audit}.

To address these limitations, we introduce an efficient end-to-end text-guided audio editing framework based on rectified flow matching (RFM) \cite{liu2022flow}, dubbed RFM-Editing. It adopts a training paradigm that learns localized velocity fields directly from instructions rather than explicit masks or captions. To support training, we construct a large-scale audio editing dataset with overlapping multi-event audio from AudioCaps2 \cite{kim-NAACL-HLT-2019}. RFM-Editing achieves competitive performance and distributional consistency across add, remove, and replace scenarios without costly inference-time optimization, even under complex overlapping events.

\vspace{-2mm}
\section{Proposed Method}
\vspace{-2mm}
\label{sec:format}

\begin{figure}[t]
\centering
\includegraphics[width=1\linewidth]{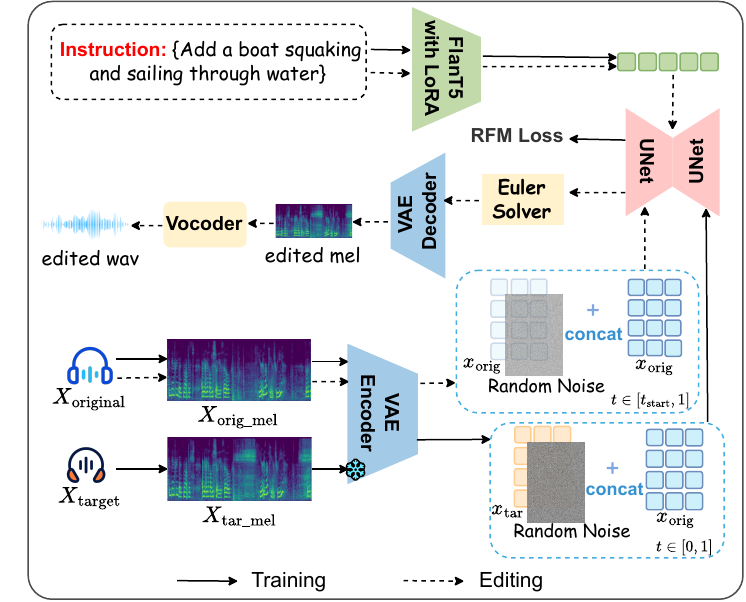}  
\caption{The training and editing pipeline of RFM-Editing.}
\label{fig:pipeline}
\vspace{-5mm}
\end{figure}

Fig.~\ref{fig:pipeline} shows the training and inference-time editing pipeline of RFM-Editing, the first unified RFM-based instruction-guided audio editing model that jointly trains three editing tasks. Built upon the LDM \cite{rombach2022high}, RFM-Editing integrates an audio feature extractor, a low-rank adaptation (LoRA \cite{hu2022lora})-tuned text encoder for instruction understanding, a U-Net for text-guided latent editing, and a BigVGAN vocoder \cite{lee2022bigvgan} for waveform reconstruction. RFM-Editing also concatenates latent with original features channel-wise and resets reverse diffusion initialization state to preserve unedited regions.

\vspace{-3mm}
\subsection{Training with Rectified Flow Matching}
RFM-Editing converts input audio to a target version based on an instruction. It is trained on original audio clips $X_{\text{original}}$, their edited counterparts $X_{\text{target}}$, and the corresponding instructions $\mathcal{I}$ by optimizing the rectified flow matching objective. Both $X_{\text{original}}$ and $X_{\text{target}}$ are converted into log-mel spectrograms $X_{\text{orig\_mel}}\in\mathbb{R}^{T \times F}$ and $X_{\text{tar\_mel}}\in\mathbb{R}^{T \times F}$, where $T$ and $F$ denote the dimensions of time and frequency, respectively. A pre-trained Variational Auto-Encoder (VAE) \cite{kingma2013auto} encodes the spectrograms into latent space, where $x_{\text{orig}}$ is the latent representation of the original audio and $x_{\text{tar}}$ that of the target audio. To better capture editing instructions, we apply LoRA \cite{hu2022lora} to the Flan-T5 text encoder \cite{chung2024scaling}, freezing pre-trained weights and inserting trainable low-rank matrices into transformer layers, reducing the number of trainable parameters while improving text understanding for precise edits.

We apply the objective of RFM \cite{liu2022flow} to learn a continuous vector field that maps samples from a noisy distribution to the target distribution of the edited audio using a U-Net-based architecture. Compared to standard diffusion models relying on stochastic differential equations (SDEs), RFM formulates a deterministic ordinary differential equation (ODE) process that models a straight-line trajectory from noise $\epsilon$ to target $x_{\text{tar}}$, eliminating the need for fine-grained time discretization and leading to stable and efficient training. Specifically, we add random Gaussian noise $\epsilon$ to $x_{\text{tar}}$ and obtain a perturbed latent $x_t$ in a continuous time step $t \in [0, 1]$. The perturbed sample $x_t$ is computed along a straight interpolation path from noise to data:
\vspace{-1.5mm}
\begin{equation}
    x_t = \left(1 - (1 - \sigma_{\min}) \cdot t\right) \cdot \epsilon + t \cdot x_{\text{tar}}
\label{addnoise}
\end{equation}
where $\sigma_{\min}$ is a small constant controlling the minimal scale of noise at $t = 0$. The time derivative of the interpolation path yields the ground-truth velocity field at any time step $t$:
\vspace{-1mm}
\begin{equation}
\mathbf{v}_{\text{target}} = \frac{d x_t}{d t} = x_{\text{tar}} - (1 - \sigma_{\min}) \cdot \epsilon
\label{eq:target}
\end{equation}

To help the model distinguish between editable and non-editable content, we provide the original latent $x_{\text{orig}}$ as an additional condition by concatenating it with the noisy latent $x_t$ along the channel dimension (as illustrated in Fig.~\ref{fig:pipeline}). This enables the model to directly access the unedited input during both training and inference, helping preserve the unchanged regions while only applying edits where instructed. RFM-Editing learns a continuous vector field $\mathbf{v}^*_\theta(x_t \oplus x_{\text{orig}}, t, E_{\mathcal{I}})$ that predicts the direction from $x_t$ to the target latent $x_{\text{tar}}$, conditioned on the instruction embedding $E_{\mathcal{I}}$. The model is trained by minimizing a mean squared error (MSE) loss between the predicted and target velocity fields:
\begin{equation}
\mathcal{L}_{\text{RFM}} = \mathbb{E}_{x_{\text{tar}}, x_{\text{orig}}, t, \epsilon} \left[ \left\| \mathbf{v}^*_\theta(x_t \oplus x_{\text{orig}}, t, E_\mathcal{I}) - \mathbf{v}_{\text{target}} \right\|^2_2 \right]
\end{equation}
This loss guides the optimization of the model parameters by encouraging the predicted vector field to align with the target velocity at each sampled timestep.

\vspace{-3mm}
\subsection{Instruction-Driven Editing}
At inference, we leverage the trained model to perform audio editing conditioned on the original audio $X_{\text{original}}$ and a textual instruction $\mathcal{I}$, without requiring the full target description, as done in \cite{jia2025audioeditor, manor2024zero}. $X_{\text{orig\_mel}}$ is first encoded by the VAE into a latent representation $x_{\text{orig}}$, while the instruction is embedded into a vector $E_{\mathcal{I}}$ using the LoRA-tuned Flan-T5 encoder.

Instead of initializing the sampling process from pure Gaussian noise, we adopt a more flexible initialization strategy inspired by DDPM/DDIM inversion. Since audio editing aims to preserve most of the original content rather than synthesizing entirely new audio from scratch, the initial state should retain partial information from the original input to better preserve unedited regions. Specifically, we define the starting point $x_{\text{start}}$ along the rectified interpolation path from noise $\epsilon$ to the original audio latent $x_{\text{orig}}$:
\begin{equation}
x_{\text{start}} = \left(1 - (1 - \sigma_{\min}) \cdot t_{\text{start}}\right) \cdot \epsilon + t_{\text{start}} \cdot x_{\text{orig}}
\label{eq:addnoise_start}
\end{equation}
where $t_{\text{start}}$ is a small adjustable parameter and we set $t_{\text{start}}=0.01$ in our model. This facilitates faithful editing by preserving non-editable regions during denoising, resulting in better consistency between the edited and original audio. Thus, the sampling interval becomes $t \in [t_{\text{start}}, 1]$. At each step $t$, the noisy latent $x_t$ is concatenated with the original latent $x_{\text{orig}}$ along the channel dimension and passed to the trained U-Net, along with the current time step $t$ and instruction embedding $E_{\mathcal{I}}$. The U-Net predicts the instantaneous velocity field $\mathbf{v}_\theta^*(x_t \oplus x_{\text{orig}}, t, E_\mathcal{I})$, which is used by a continuous-time Euler solver to iteratively update the latent:
\begin{equation}
x_{t + \Delta t} = x_t + \Delta t \cdot \mathbf{v}^*_\theta(x_t \oplus x_{\text{orig}}, t, E_\mathcal{I}).
\label{eq:euler}
\end{equation}
Iterating this update until $t=1$, we obtain the target latent $x_{\text{tar}}^*$. Finally, $x_{\text{tar}}^*$ is decoded by the VAE decoder to reconstruct the log-mel spectrogram of the edited audio. The vocoder \cite{lee2022bigvgan} is then used to convert the spectrogram into a waveform, producing the final edited audio output.

\vspace{-3mm}
\section{Experiments}
\label{sec:pagestyle}
\vspace{-3mm}
\subsection{Datasets}
\vspace{-1mm}
We construct an instruction-based audio editing dataset using AudioCaps2 \cite{kim-NAACL-HLT-2019}. The DeepSeek API is used to count sound events in each caption. Audio clips with more than three events are excluded, as they tend to be noisy and less suitable for training, and those containing only one event as single-event clips for composition. We mix each audio $X$ with two random single-event clips $A$ and $B$ to create overlapping and semantically meaningful examples, yielding six instruction-conditioned triplets: $\langle X, X+A, \text{Add A} \rangle$, $\langle X, X+B, \text{Add B} \rangle$, $\langle X+A, X, \text{Remove A} \rangle$, $\langle X+B, X, \text{Remove B} \rangle$, $\langle X+A, X+B, \text{Replace A with B} \rangle$, $\langle X+B, X+A, \text{Replace B with A} \rangle$, which are used as model inputs during training. Each example has a target caption used only for evaluation. The original captions are retained for future research.

To ensure high-quality supervision, we compute the CLAP similarity \cite{elizalde2023clap} between each audio and its caption, and retain only samples where both the original and edited pairs achieve a CLAP similarity above 0.35. The final full dataset contains 95{,}616 samples per task type, yielding 234{,}639, 26{,}103, and 26{,}103 samples for training, validation, and testing, with each split balanced across task types. We also provide a relatively smaller subset with 54{,}123, 6{,}021, and 6{,}021 samples for training, validation, and testing.

\vspace{-3mm}
\subsection{Experimental Settings and Baselines}
\vspace{-1mm}
We train the model on log-mel spectrograms with 1024 time frames and 64 mel-frequency bins extracted from 10 seconds audio clips sampled at 16kHz. The model uses a U-Net backbone with cross-attention to Flan-T5 text encoder, and is conditioned via classifier-free guidance. We adopt a velocity-based rectified flow with linear noise–data interpolation. Training is conducted for 100 epochs on A100 GPUs with a learning rate of $5\times10^{-5}$. During inference, Euler integration \cite{song2020score} is used with 200 sampling steps. Validation is based on CLAP similarity of 1000 randomly selected validation samples at each epoch during training, and the best checkpoint is saved according to the highest CLAP score.

We compare RFM-Editing against three baselines: AudioEditor \cite{jia2025audioeditor}, Zero-Shot \cite{manor2024zero} and AUDIT \cite{wang2023audit}. RFM-Editing leverages velocity-based RFM to achieve text-guided edits across diverse tasks. We use a pre-trained CLAP \cite{elizalde2023clap} to compute the cosine similarity between the edited audio and the target caption to measure semantic alignment. To assess overall quality, distributional consistency, and efficiency, we report Frechet Distance (FD), Frechet Audio Distance (FAD), Kullback–Leibler (KL) divergence, Inception Score (IS) \cite{liu2023audioldm}, and the average editing time for each audio clip. FD, KL and IS are computed using PANNs \cite{kong2020panns} that extracts both semantic embeddings and class logits, while FAD is measured with VGGish \cite{hershey2017cnn}, which captures low-level perceptual audio features. RFM-Editing refers to training on the subset, while RFM-Editing$_{\text{full}}$ denotes training on the full dataset.

\vspace{-4mm}
\subsection{Results}
\vspace{-1mm}
\begin{table}[t]
\centering
\footnotesize
\renewcommand{\arraystretch}{1.00}
\setlength{\abovecaptionskip}{4pt}
\caption{Quantitative evaluation of edited audio.}
\label{table1Quantitative}
\begin{tabular}{lcccc}
\toprule
\textbf{Method} & \textbf{FD ↓} & \textbf{FAD ↓} & \textbf{KL ↓} & \textbf{IS ↑} \\ 
\midrule
AudioEditor \cite{jia2025audioeditor}           & \underline{14.24} & \textbf{2.01} & 4.07          & \textbf{8.40} \\ 
AUDIT \cite{wang2023audit}                & 32.62          & 7.22          & 9.99          & \underline{6.59} \\ 
Zero-Shot \cite{manor2024zero}             & 25.77          & 3.86          & 4.09          & 5.04          \\ 
\midrule
RFM-Editing            & 15.00          & 2.95          & \underline{2.90} & 4.90          \\ 
RFM-Editing$_{\text{full}}$ & \textbf{13.27} & \underline{2.50} & \textbf{2.77} & 5.27          \\ 
\bottomrule
\end{tabular}
\vspace{-4mm}
\end{table}

\begin{figure*}[t]
\centering
\begin{minipage}[b]{0.49\linewidth}
  \centering
  \includegraphics[width=0.9\linewidth]{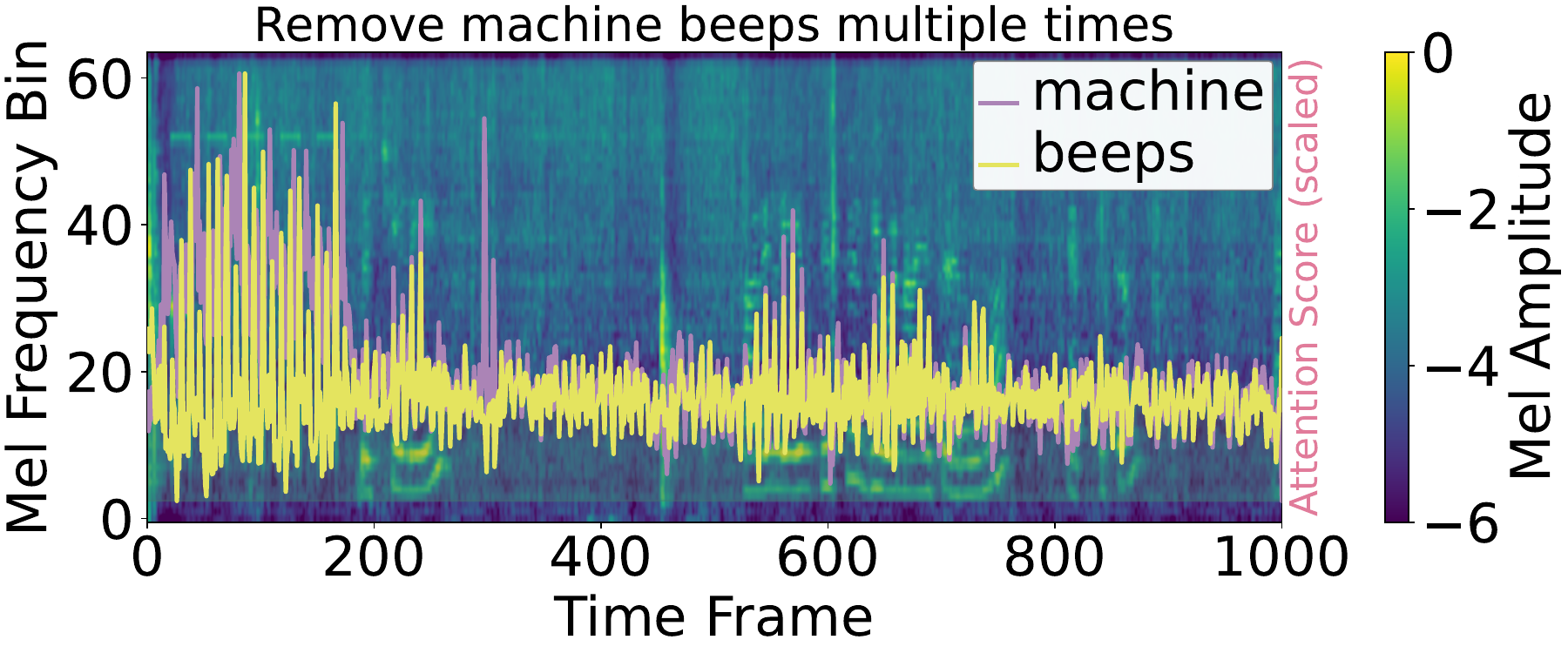}
\end{minipage}
\hfill
\begin{minipage}[b]{0.49\linewidth}
  \centering
  \includegraphics[width=0.9\linewidth]{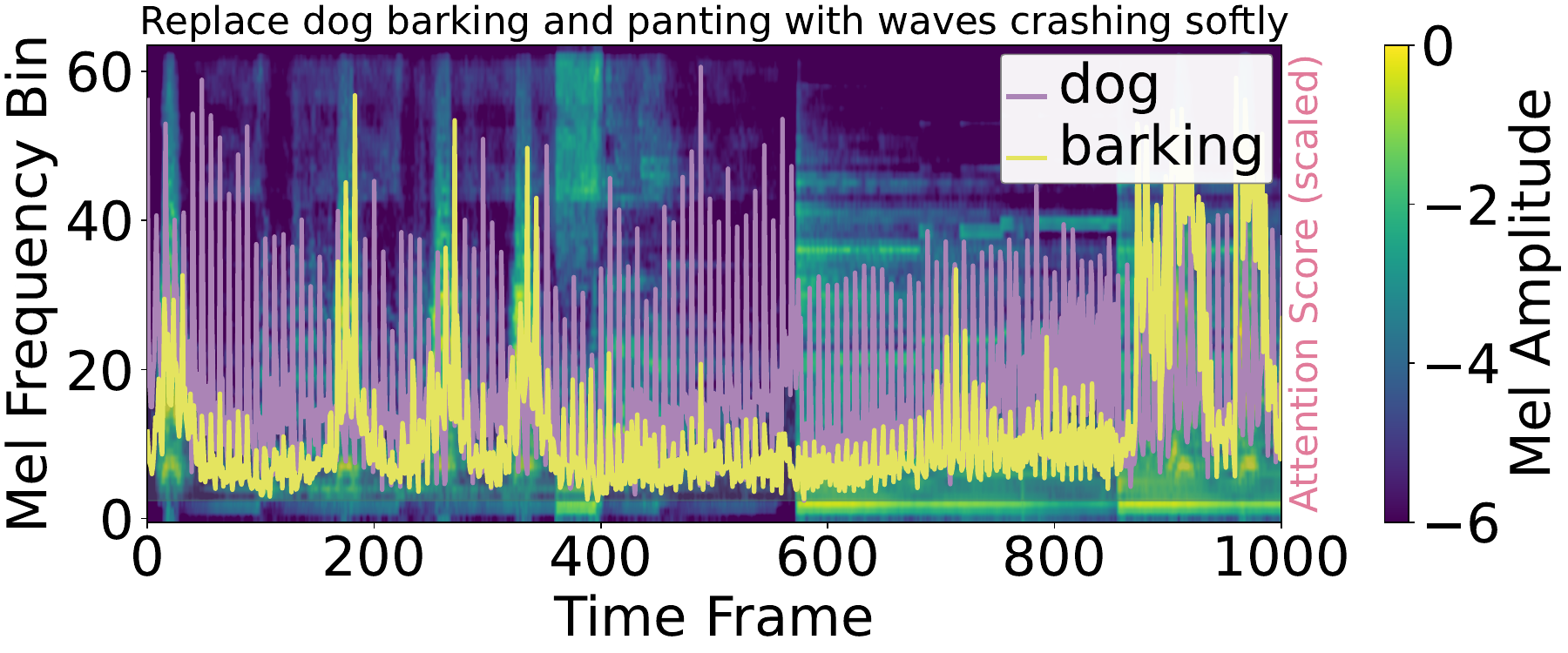}
\end{minipage}
\captionsetup{skip=1pt}
\caption{Visualizations of dynamic cross-attention trajectories of specific tokens in remove and replace tasks in RFM-Editing.}
\label{fig2:dynamicattention}
\vspace{-3mm}
\end{figure*}

\captionsetup{skip=1pt}  
\setlength{\abovecaptionskip}{6pt} 
\setlength{\belowcaptionskip}{0pt}

\begin{figure*}[t]
\centering
\begin{minipage}[b]{0.32\linewidth}
  \centering
  \includegraphics[width=\linewidth]{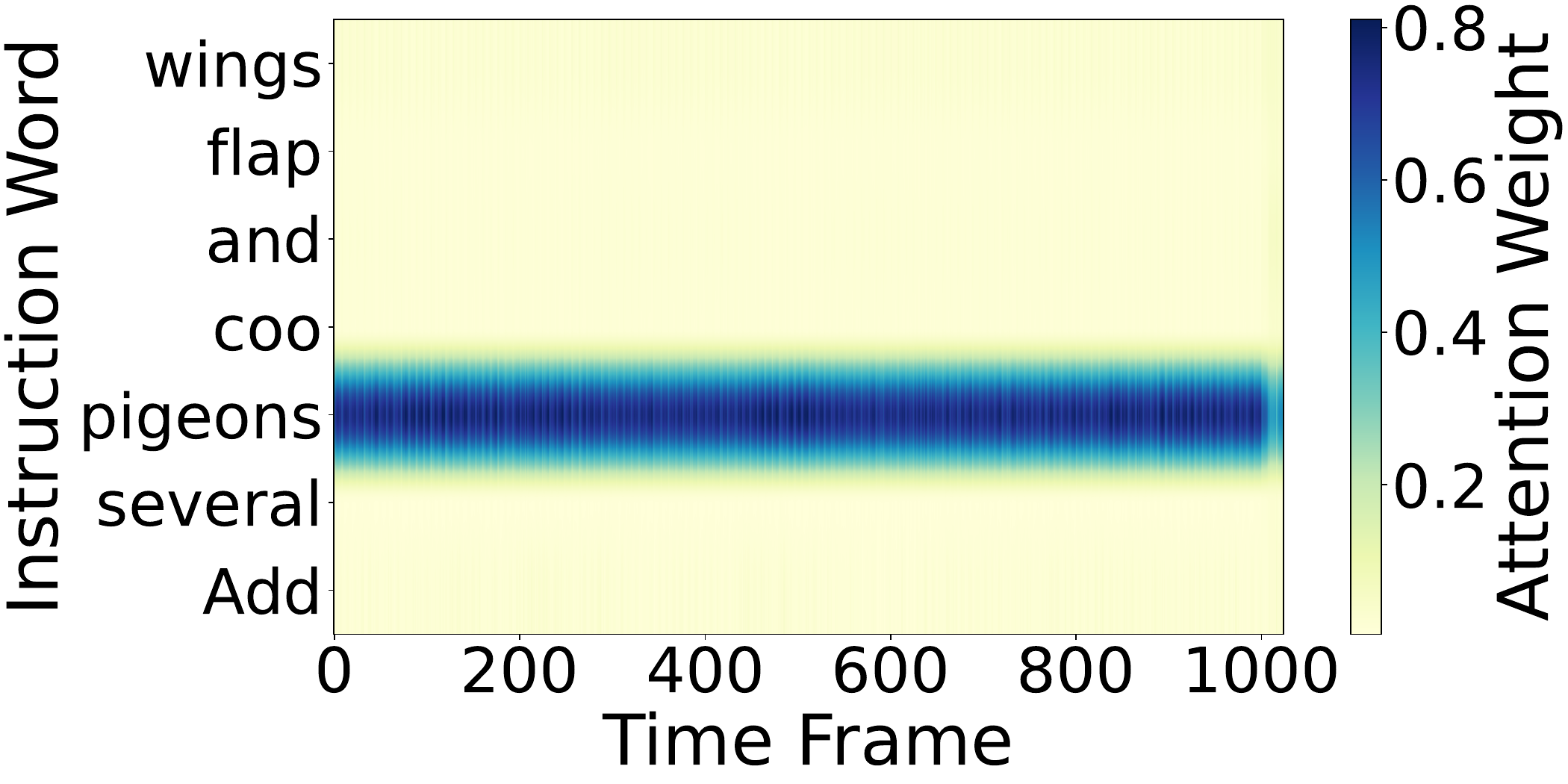}
  \captionsetup{skip=1pt} 
  \caption*{\small (a) Add}
\end{minipage}
\hfill
\begin{minipage}[b]{0.32\linewidth}
  \centering
  \includegraphics[width=\linewidth]{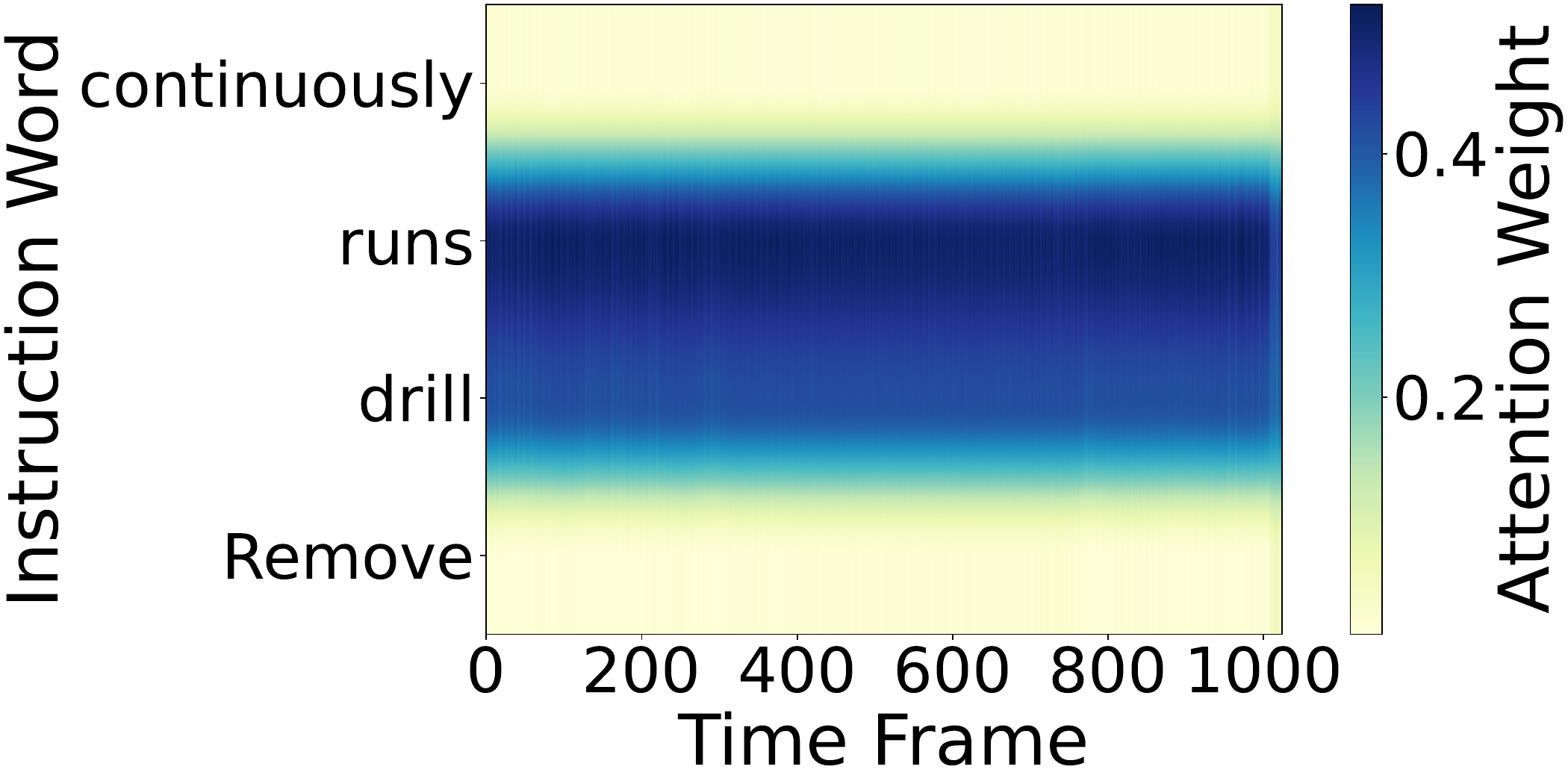}
  \captionsetup{skip=1pt}
  \caption*{\small (b) Remove}
\end{minipage}
\hfill
\begin{minipage}[b]{0.32\linewidth}
  \centering
  \includegraphics[width=\linewidth]{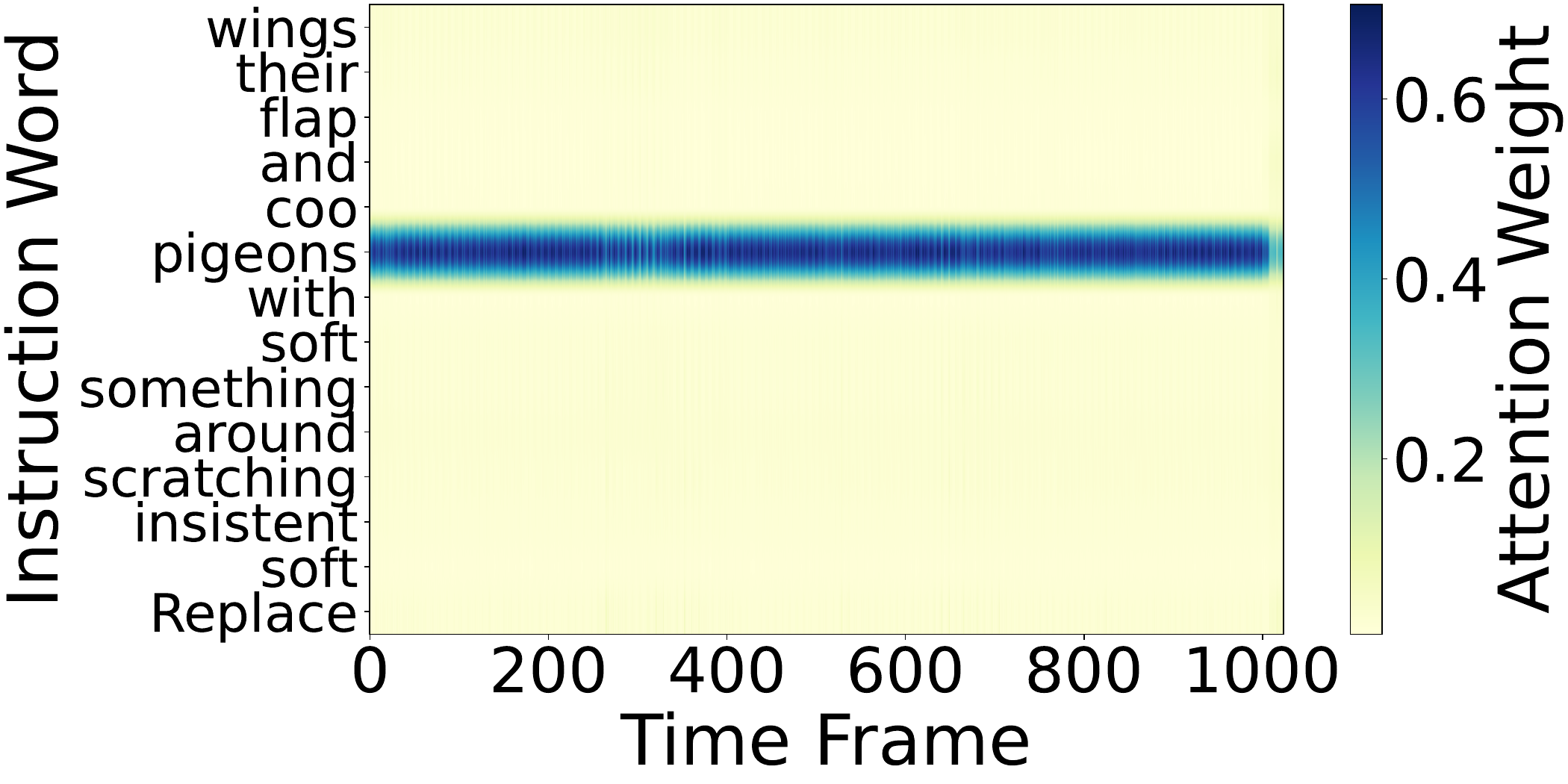}
  \captionsetup{skip=1pt}
  \caption*{\small (c) Replace}
\end{minipage}
\caption{Visualization of token-wise cross-attention distributions between the instruction sequence and audio features.}
\vspace{-5mm}
\label{fig3:heatmap}
\end{figure*}

Table~\ref{table1Quantitative} reports the quantitative comparison in terms of edited audio quality, including FD, FAD, KL, and IS. We observe that RFM-Editing trained with the subset already achieves competitive performance, substantially outperforming AUDIT \cite{wang2023audit} and the Zero-Shot \cite{manor2024zero} across most metrics. When trained on the full dataset, RFM-Editing$_{\text{full}}$ further improves and achieves the best FD and KL scores, indicating higher distributional consistency and better alignment with the target semantics distribution in feature space. These results suggest that velocity-based RFM enables more stable modeling of global distributional transitions and stronger generalization to diverse semantic shifts specified by instructions in audio editing. Although RFM-Editing obtains only moderate IS, this is expected as the task emphasizes faithful alignment with editing instructions rather than maximizing output diversity.

While AudioEditor \cite{jia2025audioeditor} attains the lowest FAD by local attention manipulation, its performance on global semantic consistency shown by KL is limited due to over-editing. AUDIT \cite{wang2023audit} suffers from poor fidelity and distributional scores after being trained for 100 epochs on the constructed dataset, whereas Zero-Shot \cite{manor2024zero} achieves competitive performance but lacks distributional consistency. In contrast, our approach achieves a more balanced performance across all metrics. 

\begin{table}[t]
\centering
\footnotesize
\setlength{\tabcolsep}{5pt}
\renewcommand{\arraystretch}{1.00}
\caption{Comparison of editing fidelity and efficiency.}
\label{tab:my-table}
\begin{tabular}{lccc}
\toprule
\textbf{Method} & \textbf{Prompt} & \textbf{CLAP ↑} & \textbf{Editing Time (s) ↓} \\ 
\midrule
AudioEditor \cite{jia2025audioeditor} & \begin{tabular}[c]{@{}c@{}}caption \&\\ modified tokens\end{tabular} & \textbf{0.4579} & 101.87 \\
AUDIT \cite{wang2023audit} & \textbf{instruction} & 0.1113 & \underline{11.00} \\
Zero-Shot \cite{manor2024zero} & \underline{caption} & 0.4333 & 12.52 \\
\midrule
RFM-Editing & \textbf{instruction} & 0.4250 & \textbf{10.97} \\
RFM-Editing$_{\text{full}}$ & \textbf{instruction} & \underline{0.4398} & 11.27 \\
\bottomrule
\end{tabular}
\vspace{-5mm}
\end{table}

In Table~\ref{tab:my-table}, RFM-Editing demonstrates a clear advantage in faithfulness of the edited audio to the target captions and the editing efficiency. AudioEditor \cite{jia2025audioeditor} achieves the best CLAP score and alignment with the target text due to attention replacement mechanism \cite{hertzprompt}, but its editing is nearly an order of magnitude slower than ours due to inference-time optimization, significantly degrading user experience. Moreover, it requires both full captions and modified token indices and the Zero-Shot \cite{manor2024zero} depends on target captions rather than concise editing instructions, which highlights the superiority of instruction-driven methods such as \cite{wang2023audit} and RFM-Editing.

\vspace{-4mm}
\subsection{Ablation and Visualization}
\vspace{-2mm}
We analyze the effect of initialization for diffusion start time on editing behavior in Table~\ref{table3}. Increasing $t_\text{start}$ preserves more of the original audio, but reduces the editing strength, which results in improved perceptual quality, as reflected by the lowest FAD and highest IS when $t_\text{start}=0.1$, but simultaneously leads to poor semantic alignment with the target captions, as indicated by an extremely low CLAP score. By contrast, setting $t_\text{start}=0.01$ offers the best trade-off, yielding the best CLAP while still maintaining competitive audio quality.

To intuitively illustrate the effectiveness of RFM-Editing, we visualize the cross-attention weights within the diffusion network. Successful editing requires the model to accurately localize the time frames of sound events to be removed or replaced, while also attending to newly added events in the instruction. We present two complementary visualizations: dynamic normalized cross-attention trajectories between selected instruction tokens and audio features in Fig.~\ref{fig2:dynamicattention}, and token-wise cross-attention heatmaps that illustrate the relative attention distribution of different tokens throughout the generation process in Fig.~\ref{fig3:heatmap}.

In Fig.~\ref{fig2:dynamicattention}, the cross-attention weights between specific tokens and audio features reveal how the model focuses on key segments for editing. Tokens ``beeps" and ``barking" associated with transient events exhibit sharp attention peaks that align with actual sound occurrences, whereas the token ``dog" describing a persistent source maintains high attention over longer time spans. This observation reflects the true temporal structure of the audio and indicates that the model can accurately localize target events without time-aligned masks, which is crucial for precise and effective audio editing.
\begin{table}[t]
\centering
\footnotesize
\renewcommand{\arraystretch}{1.00}
\caption{Effect of diffusion start time initialization.}
\label{table3}
\begin{tabular}{cccccc}
\toprule
\textbf{$t_\text{start}$} & \textbf{CLAP ↑} & \textbf{FD ↓} & \textbf{FAD ↓} & \textbf{KL ↓} & \textbf{IS ↑} \\ 
\midrule
0     & 0.4216 & 17.97 & \underline{2.45} & \underline{2.96} & 4.27 \\ 
0.001 & \underline{0.4224} & 17.94 & 2.48 & \textbf{2.94} & 4.27 \\ 
0.01  & \textbf{0.4249} & \underline{17.38} & 2.52 & 3.06 & \underline{4.34} \\ 
0.1   & 0.3799 & \textbf{16.80} & \textbf{1.49} & 4.47 & \textbf{5.24} \\ 
\bottomrule
\end{tabular}
\vspace{-5mm}
\end{table}

Furthermore, the heatmaps in Fig.~\ref{fig3:heatmap} (a), (b) and (c) show that the model consistently attends to the key parts of the instruction across all tasks, ensuring accurate and instruction-aligned editing outcomes. Interestingly, we observe that in replacement tasks, if the model assigns greater attention to the events to be removed rather than the newly introduced ones, the editing quality tends to degrade. In addition, the quality of the instruction has a substantial impact on the results, highlighting the critical role of prompts in editing tasks.

\vspace{-3.8mm}
\section{Conclusion}
\vspace{-2.3mm}
\label{sec:typestyle}
We have presented RFM-Editing, the first rectified flow matching framework for instruction-guided audio editing without captions or masks, along with a new dataset. Experiments show that RFM-Editing can automatically localize instruction-relevant time frames, achieving faithful alignment with target semantics and precise editing. Results highlight rectified flow matching as a practical paradigm, and suggest future work on leveraging language prompting capabilities.

\newpage

\section{Acknowledgment}
{\small This work was partially supported by a research scholarship from the China Scholarship Council (CSC) and a studentship from the University of Surrey. For the purpose of open access, the authors have applied a Creative Commons Attribution (CC BY) license to any Author Accepted Manuscript version arising from this work.}

\normalsize
\section{REFERENCES}

\begingroup
\small
\bibliographystyle{IEEEtran}
\bibliography{refs}
\endgroup

\end{document}